\documentclass[manuscript,screen,nonacm]{acmart}

\usepackage[T1]{fontenc}
\usepackage[utf8]{inputenc}       

\usepackage{graphicx}
\usepackage{subcaption}
\usepackage[dvipsnames,table]{xcolor}
\definecolor{aceBlue}{HTML}{2166AC} 
\definecolor{customgreen1}{RGB}{119,221,119} %
\definecolor{customred1}{RGB}{221,119,119}

\usepackage{booktabs}             
\usepackage{multirow}             
\usepackage{array}                
\usepackage{makecell}             
\usepackage{xspace} 
\usepackage{mdframed}
\usepackage{ragged2e} 
\usepackage{tabularx} 

\usepackage{microtype}            
\usepackage{relsize}              

\usepackage{enumitem}

\usepackage{hyperref}

\newcommand{\textcode}[1]{{\fontfamily{cmtt}\selectfont #1}\xspace}

\newcommand{\move}{\texttt{move}}
\newcommand{\turnL}{\texttt{turnLeft}}
\newcommand{\turnR}{\texttt{turnRight}}

\newcommand{\DSLRepeat}{\textcode{\textsc{Repeat}}}
\newcommand{\DSLRepeatUntil}{\textcode{\textsc{RepeatUntil}}}
\newcommand{\DSLIf}{\textcode{\textsc{If}}}
\newcommand{\DSLIfElse}{\textcode{\textsc{IfElse}}}

\newcommand{\nohint}{\textsc{None}}
\newcommand{\codeedit}{\textsc{Code-Rec}}
\newcommand{\solcode}{$C^{\ast}$}
\newcommand{\stucode}{$C^\text{\small{stu}}$}
\newcommand{\codehint}{$C^\text{\small{rec}}$}
\newcommand{\quizadapt}{$Q^\text{\small{rec}}$}
\newcommand{\quizplan}{$Q^\text{\small{plan}}$}
\newcommand{\quizsolve}{$Q^\text{\small{solve}}$}
\newcommand{\adaptquiz}{\textsc{Code-Quiz}}
\newcommand{\planquiz}{\textsc{Plan-Quiz}}

\newcommand{\hoc}{\textsc{HoC}}
\newcommand{\posttest}{\textsc{PostHoC}}
\newcommand{\learningeasy}{\textsc{Easy}\ensuremath{_\texttt{L}}}
\newcommand{\learninghard}{\textsc{Hard}\ensuremath{_\texttt{L}}}
\newcommand{\postlearningeasy}{\textsc{Easy}\ensuremath{_\texttt{PL}}}
\newcommand{\postlearninghard}{\textsc{Hard}\ensuremath{_\texttt{PL}}}
\newcommand{\postlearningnew}{\textsc{New}\ensuremath{_\texttt{PL}}}
\newcommand{\postlearningcommon}{\textsc{Common}\ensuremath{_\texttt{PL}}}

\newcommand{\lowexpstudent}{\textsc{LowYoS}}
\newcommand{\highexpstudent}{\textsc{HighYoS}}

\begin{document}

\title[Effectiveness of Feedback Methods in  Elementary-Level Visual Programming]{The Right Kind of Help: Evaluating the Effectiveness of  Feedback Methods in  Elementary-Level Visual Programming}
\titlenote{Preprint. Accepted as a paper at the ISSEP'26 conference.}

\author{Ahana Ghosh}
\orcid{0000-0002-0967-5886}
\affiliation{%
  \institution{Max Planck Institute for Software Systems}
  \city{Saarbrücken}
  \country{Germany}
}
\email{gahana@mpi-sws.org}

\author{Liina Malva}
\orcid{0000-0002-9731-8247}
\affiliation{%
  \institution{Max Planck Institute for Software Systems}
  \city{Saarbrücken}
  \country{Germany}
}
\affiliation{%
  \institution{Tallinn University}
  \city{Tallinn}
  \country{Estonia}
}
\email{liina.malva@tlu.ee}

\author{Alkis Gotovos}
\orcid{0000-0002-3902-8890}
\affiliation{%
  \institution{Max Planck Institute for Software Systems}
  \city{Saarbrücken}
  \country{Germany}
}
\email{agkotovo@mpi-sws.org}

\author{Danial Hooshyar}
\orcid{0000-0002-9143-6648}
\affiliation{%
  \institution{Tallinn University}
  \city{Tallinn}
  \country{Estonia}
}
\email{danial.hooshyar@tlu.ee}

\author{Adish Singla}
\orcid{0000-0001-9922-0668}
\affiliation{%
  \institution{Max Planck Institute for Software Systems}
  \city{Saarbrücken}
  \country{Germany}
}
\email{adishs@mpi-sws.org}

\setcounter{footnote}{0}

\begin{abstract}
	\looseness-1We present a large-scale study comparing the effectiveness of various feedback methods in elementary-level programming. While prior work has explored different feedback methods, their relative impact during the learning and post-learning phases remains unclear.
    In this study, we compare three feedback methods: code-edit recommendations (\codeedit), quizzes based on code edits (\adaptquiz), and quizzes based on metacognitive strategies (\planquiz), along with a no-feedback control (\nohint). A total of $398$ students (across grades $4$–$7$) participated in a two-phase study: a learning phase comprising write-code tasks from the \emph{Hour of Code: Maze Challenge} with feedback, followed by a post-learning phase comprising more advanced write-code tasks without feedback.
    All feedback methods significantly improved learning performance over the control, while preserving students' problem-solving skills in the post-learning phase. Furthermore, quiz-based methods showed a consistent trend of stronger performance on novel post-learning tasks. Students in feedback groups also reported greater engagement and perceived skill growth.
\end{abstract}


\maketitle              


\section{Introduction}\label{sec:intro}

\looseness-1Block-based visual programming environments have become an integral part of K-$8$ programming education worldwide via widely adopted platforms such as the \emph{Hour of Code: Maze Challenge}~\cite{hourofcode_maze} by code.org~\cite{codeorg}, \emph{Scratch}~\cite{DBLP:journals/jeric/MaloneyRRSE10}, and Stanford's \emph{Programming with Karel the Robot}~\cite{StanfordKarel,Pattis1994KarelTR}. However, programming tasks on these platforms can be inherently conceptual and open-ended, requiring multi-step deductive reasoning, making them challenging for students~\cite{DBLP:conf/sigcse/GroverBS18,DBLP:conf/lasi-spain/KanellopoulouGG21,DBLP:conf/lats/PiechSHG15}. Hence, to effectively support a struggling student in solving these tasks, it is important to design just-in-time feedback on their attempts~\cite{doi:10.3102/0034654314564881}.

Human tutors in programming education typically design feedback to be both task-specific and generalizable, ensuring that feedback not only corrects immediate errors but also guide learners' thinking, reinforce strategy use, and cultivate self-regulated learning (SRL) behaviors such as planning, monitoring, and reflection~\cite{Nicol01042006,Zimmerman01052002}. Moreover, passive feedback that reveals solutions could impede metacognitive processes, limiting learners' ability to transfer knowledge to new contexts~\cite{Chi02102014,CHI1989145}. Educational frameworks such as Cognitive Load Theory~\cite{SWELLER1994295} and SRL suggest that feedback must balance reducing unnecessary complexity with preserving opportunities for active engagement~\cite{doi:10.3102/00346543065003245}, while also supporting learners' sense of autonomy, competence, and skill development~\cite{DBLP:books/sp/DeciR85}.

\begin{figure*}[!t]
\centering
        \centering
        \includegraphics[width=0.95\textwidth]{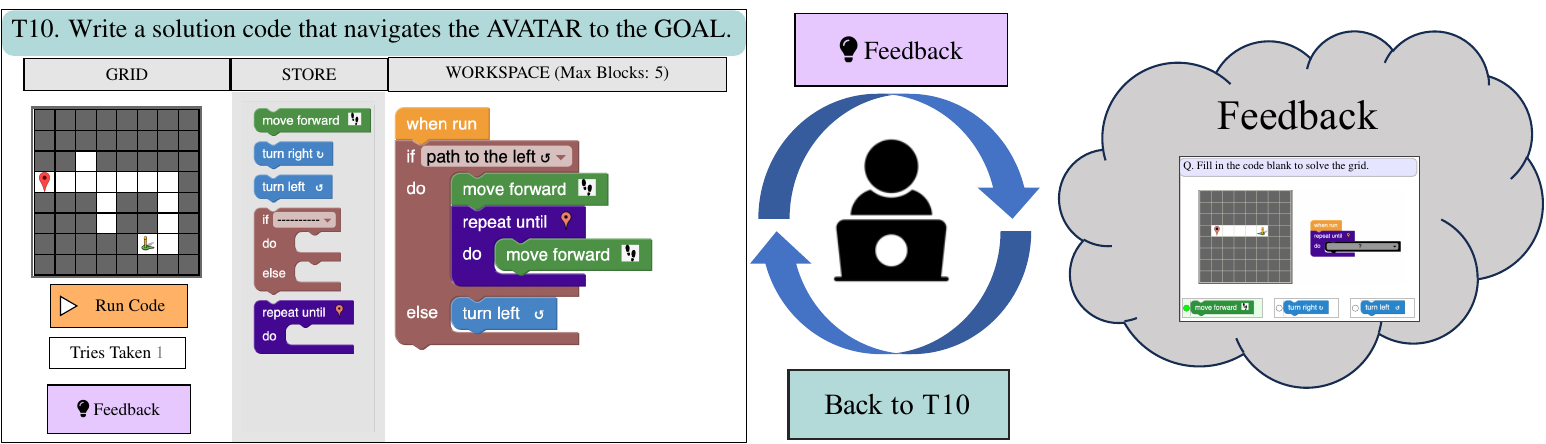}
    \caption{Illustration of our framework for providing feedback. The interface shows a student working on task T10 during the learning phase. The student can seek feedback at any time via the ``Feedback'' button, which links to a specific method. After interacting with the feedback, the student can continue working on the task. The no-feedback control (\nohint{}) does not have this ``Feedback'' button.}
    \label{fig1:intro}
\end{figure*}

\looseness-1Prior work in programming education has developed automated feedback methods including next-step code edits~\cite{Paassen_Hammer_Price_Barnes_Gross_Pinkwart_2018}, worked examples~\cite{DBLP:conf/sigcse/ZhiPMMBC19}, code-level quizzes~\cite{DBLP:conf/aied/GhoshTDS22}, and metacognitive scaffolds~\cite{DBLP:conf/ace/PechorinaAD23}. These methods have been primarily evaluated for their effectiveness during learning. However, their impact on promoting independent problem-solving in the absence of feedback post-learning has not been as extensively explored. In particular, whether these methods induce overreliance when students apply learned skills to new tasks remains an open question. In this work we address this gap by investigating the following key question: \emph{How effective are feedback methods in elementary-level visual programming for supporting students during learning and fostering independent problem-solving?}

\looseness-1To investigate this, we designed a large-scale study with elementary school students comparing three feedback methods: (i) code recommendations based on next-step code edits, (ii) quizzes based on recommended code edits, and (iii) quizzes based on metacognitive problem-solving stages such as task planning and solution finding. A no-feedback control group serves as a baseline. These methods were evaluated across a learning phase (with feedback) and a post-learning phase (without feedback). Figure~\ref{fig1:intro} illustrates the framework for providing feedback. Our analysis addresses: \textbf{(RQ1)} How effective are the feedback methods in supporting students' task performance and cognitive effort during the learning phase? \textbf{(RQ2)} How effective are the feedback methods in fostering independent problem-solving in the post-learning phase? \textbf{(RQ3)} How do students perceive the feedback methods in terms of interest, enjoyment, and skill development?


\section{Related Work}\label{sec:related_work}

\looseness-1\textbf{Design of feedback methods.} Within programming education, prior work has explored diverse feedback formats including code recommendations~\cite{Paassen_Hammer_Price_Barnes_Gross_Pinkwart_2018,DBLP:conf/lats/PiechSHG15}, worked examples~\cite{DBLP:conf/sigcse/ZhiPMMBC19}, scaffolds~\cite{DBLP:conf/ace/PechorinaAD23}, and textual explanations~\cite{DBLP:conf/sigcse/WangMP24}, each varying in how they engage learners' cognitive and metacognitive processes. Effective feedback should actively guide learners, encourage reflection, and be timely and task-focused to foster schema construction and transfer~\cite{Nicol01042006,doi:10.3102/0034654307313795}. However, no studies have explicitly compared these feedback modalities in elementary-level programming. We evaluate three representative feedback methods: a code-recommendation method based on next-step code edits, a fill-in-the-gap quiz adapted from the learner's attempt and inspired by worked examples, and a planning and solution-finding quiz targeting metacognitive processes.

\looseness-1\textbf{Cognitive load of feedback.} Grounded in Cognitive Load Theory (CLT)\cite{SWELLER1994295}, instructional materials must balance meaningful guidance with cognitive demands on working memory. Effective feedback reduces extraneous load while preserving opportunities for active engagement~\cite{doi:10.3102/0034654307313795,c1b289fe-5171-318a-b656-0fcdb00b93af}; excessive scaffolding or late-stage feedback can increase unnecessary load~\cite{Kalyuga01012003,Kirschner01062006}. This motivates \emph{RQ1}, examining feedback methods' effectiveness in supporting students' task performance and cognitive effort during the learning phase.

\looseness-1\textbf{Cognitive engagement and transfer effects.} Beyond immediate performance, effective feedback should foster transferable knowledge, enabling learners to apply strategies to novel problems once scaffolds are removed~\cite{c1b289fe-5171-318a-b656-0fcdb00b93af,Zimmerman01052002}. Feedback that scaffolds metacognitive monitoring or uses faded worked examples has been linked to improved far transfer in domains requiring abstraction and decomposition~\cite{RENKL2002529}. Yet in programming education, post-learning transfer effects of feedback remain understudied. We assess transfer via students' performance on (novel) post-learning tasks completed without any feedback. This motivates \emph{RQ2}, examining feedback methods' effectiveness in fostering independent problem-solving.

\looseness-1\textbf{Learner motivation.} Motivation both drives and emerges from the SRL cycle, shaped by learners' sense of autonomy, competence, and enjoyment~\cite{DBLP:books/sp/DeciR85,Zimmerman01052002}. In programming education, interactive and challenge-based activities improve affective outcomes including enjoyment and self-efficacy~\cite{DBLP:conf/icer/MarwanGFPB20,doi:10.3102/0034654307313795}, and interleaving learning with formative assessment sustains engagement~\cite{bjork2011making}. This motivates \emph{RQ3}, examining how students perceive the feedback methods in terms of interest, enjoyment, and skill development.


\begin{figure*}[!h]
\centering
    \begin{subfigure}[c]{0.47\textwidth}
       \centering
       \fbox{\includegraphics[width=0.99\textwidth]{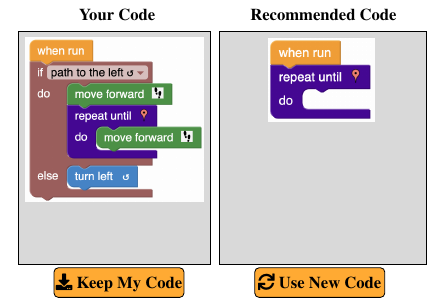}}
        \caption{Feedback method: \codeedit{}}
        \label{fig1.codeedit}
    \end{subfigure}
    \hspace{1em}
    \begin{subfigure}[c]{0.47\textwidth}
        \centering
        \fbox{\includegraphics[width=0.99\textwidth]{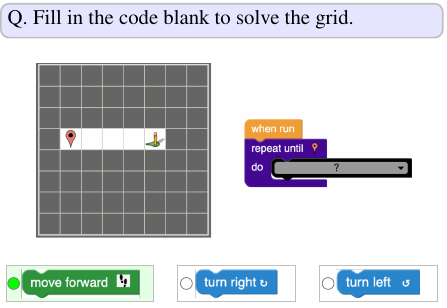}}
        \caption{Feedback method: \adaptquiz{}}
        \label{fig1.adaptquiz}
    \end{subfigure}
    \\
   \vspace{1mm}
    \begin{subfigure}[c]{\textwidth}
        \centering
        \begin{minipage}[c]{0.47\textwidth}
            \centering
            \fbox{\includegraphics[width=0.99\textwidth]{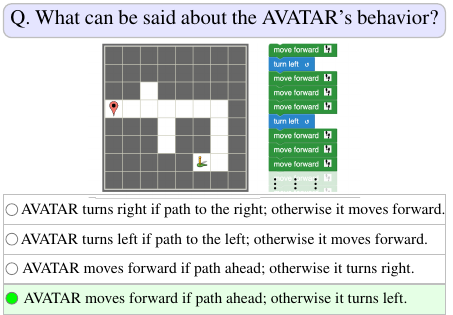}}
        \end{minipage}
        \hspace{1em}
        \begin{minipage}[c]{0.47\textwidth}
            \centering
            \fbox{\includegraphics[width=0.99\textwidth]{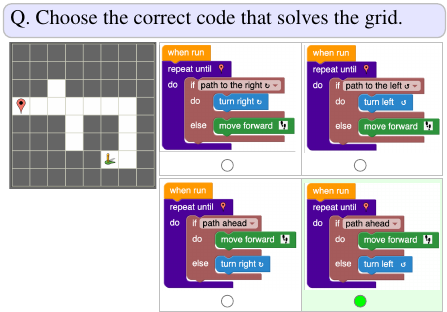}}
        \end{minipage}
        \caption{\looseness-1Feedback method: \planquiz{}. (Left) quiz \planquiz{} I is based on task planning; (right) quiz \planquiz{} II is based on solution finding.}
        \label{fig1.planquiz12}
    \end{subfigure}
    \caption{\looseness-1~Illustration of feedback methods evaluated in our study, corresponding to task T10 and student attempt shown in Figure~\ref{fig1:intro}. (a) \codeedit{} and (b) \adaptquiz{} provide feedback adapted to the student's current attempt. (c) \planquiz{} provides two quizzes for the task, not adapted to the student's current attempt.}
    \label{fig2:feedback_methods}
\end{figure*}

\section{Feedback Methods}\label{sec:method}

\looseness-1We describe the three feedback methods evaluated in our study. As shown in Figure~\ref{fig1:intro}, students can request feedback via the ``Feedback'' button at any time during the learning phase, and are prompted after three consecutive incorrect attempts. Implementation details of the methods are presented in Appendix~\ref{sec:appendix}.

\looseness-1\textbf{No feedback (\nohint{}).} This control group does not have access to feedback during the learning phase, i.e., there is no ``Feedback'' button (see Figure~\ref{fig1:intro}). It serves as a baseline for comparing the performance of other feedback methods.

\textbf{Code-edit recommendations (\codeedit{}).} This feedback method provides adaptive code recommendations to guide students toward the correct solution (\solcode) for the task. Given a student's attempt (\stucode), the method finds an intermediate code (\codehint) which brings \stucode{} structurally closer to \solcode{}. This method is inspired by next-step edits~\cite{Paassen_Hammer_Price_Barnes_Gross_Pinkwart_2018,DBLP:conf/lats/PiechSHG15,DBLP:conf/its/RiversK14}, though it makes these edits at a code structure-level instead of individual blocks. As shown in Figure~\ref{fig1.codeedit}, the feedback is provided in the form of recommendations, where students can either retain their attempt (via the ``Keep My Code'' button) or apply the recommendation (via the ``Use New Code'' button). 

\looseness-1\textbf{Quizzes based on code edits (\adaptquiz{}).} This feedback method integrates code-level feedback with interactive problem-solving, by transforming \codeedit{} recommendations into fill-in-the-gap quizzes. This method is inspired by quiz-based feedback in prior work~\cite{DBLP:conf/aied/GhoshTDS22,DBLP:conf/sigcse/ZhiPMMBC19}. As shown in Figure~\ref{fig1.adaptquiz}, each quiz has a visual grid, partial block-based code, and three options (e.g., \move, \turnL, \turnR). Students choose the correct code block to complete the blank.

\textbf{Quizzes based on metacognitive strategies (\planquiz{}).} This feedback method integrates metacognitive strategies with interactive problem-solving, by presenting non-adaptive quizzes based on task planning and solution finding. This method is inspired by the metacognitive scaffolding framework presented in the work of Pechorina et al.~\cite{DBLP:conf/ace/PechorinaAD23} and is simplified in the context of elementary programming. As shown in Figure~\ref{fig1.planquiz12}, the feedback provides two quizzes: the first quiz asks students to analyze patterns in the action sequences; the second quiz asks them to select the correct solution. Students may revisit both quizzes at any point of their problem-solving process on the task.


\section{Study Setup}\label{sec:setup}

\subsection{Learning and Post-Learning Phases}\label{sec:method.tasks}

\begin{figure*}[h]
\centering
      \includegraphics[width=\textwidth]{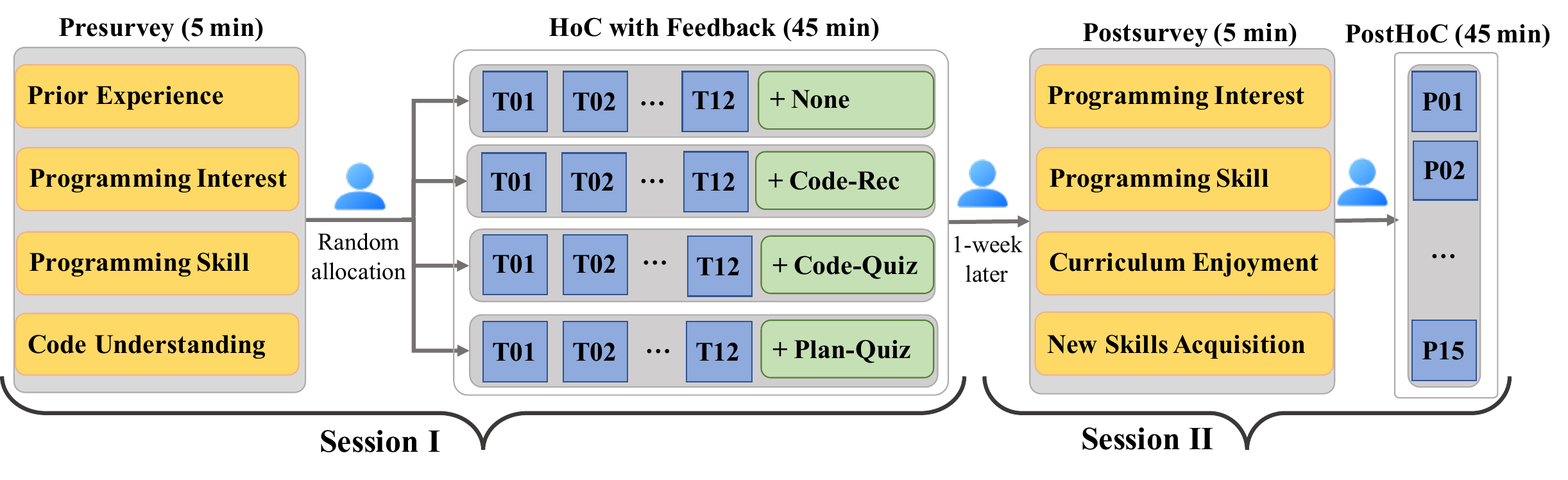}
    \caption{Illustration of our two-session study setup.}
    \label{fig:study_framework}
\end{figure*}

\looseness-1Our study comprises two sessions serving as a proxy for the learning and post-learning phases respectively; see Figure~\ref{fig:study_framework}. Session~I comprises one $50$-minute lesson, where participants begin with a brief presurvey followed by a curriculum of write-code tasks in their assigned feedback group, with each task allowing multiple attempts. Session~II also comprises one $50$-minute lesson beginning with a brief postsurvey followed by a curriculum of write-code tasks without any feedback. 

\looseness-1\textbf{Session~I curriculum.} This phase evaluates the effectiveness of feedback methods during problem-solving. Given the limited duration of $50$ minutes, and accounting for the time students would need to engage with the feedback, we selected $12$ write-code tasks after consulting with teachers. These tasks are from the popular programming curriculum, \emph{Hour of Code: Maze Challenge}~\cite{hourofcode_maze} by code.org~\cite{codeorg}, denoted as T01--T12 (collectively referred to as \hoc{}), and cover introductory programming concepts as shown in Figure~\ref{fig:dist.hocposttest}. We categorize tasks according to the complexity of their solution codes as follows: \learningeasy{} (T01--T04) include basic actions such as moves, turns, and simple loops; \learninghard{} (T05--T12) include more advanced programming structures such as loops and conditionals. Distribution of these tasks according to the programming concepts of solution codes is shown in Figure~\ref{fig:dist.hocposttest} (Appendix~\ref{sec:appendix.results}).

\looseness-1\textbf{Session~II curriculum.} This phase evaluates independent problem-solving without any feedback. Given that this phase did not involve any feedback, we used a slightly larger pool of $15$ write-code tasks based on prior work and input from the teachers~\cite{DBLP:conf/sigcse/GhoshMGHS25}. These tasks, denoted as P01--P15 (collectively referred to as \posttest{}), are also based on the same domain, though some tasks cover new combinations of concepts not present in the \emph{Hour of Code: Maze Challenge}; see Figure~\ref{fig:dist.hocposttest}. Similar to the learning phase, we categorized the tasks by difficulty: \postlearningeasy{} (P01,P02,P08); \postlearninghard{} (P03,P04--P07,P09--P15). Additionally, we characterize the tasks based on their similarity to \hoc{} as follows: \postlearningcommon{} (P01,P02, P04--P07) tasks are a subset of \hoc{}; \postlearningnew{} (P12--P14) tasks are novel to \posttest{}, requiring new combinations of programming concepts not seen in \hoc{}. Distribution of these tasks based on programming concepts and illustrations of novel write-code tasks from this phase are shown in Figures~\ref{fig:dist.hocposttest} and ~\ref{fig:tasks.example.posttest}, respectively (Appendix~\ref{sec:appendix.results}).

We use task success as a proxy for learning performance in Session~I and post-learning transfer in Session~II; time-on-task and feedback time serve as proxies for cognitive effort and engagement.

\subsection{Study Logistics and Demographics}\label{sec:setup.phases}
\looseness-1We obtained an Institutional Review Board (IRB) approval from the Ethics Committee of Tallinn University before conducting the study. Following the data collection procedures of our previous studies~\cite{DBLP:conf/sigcse/GhoshMGHS25,DBLP:conf/sigcse/GhoshMS24}, we collected data in Estonia between January and March 2025. We contacted $22$ new schools by directly reaching out to their teachers. Participation in the study was entirely voluntary for students. Participating students were randomly allocated to one of the four feedback groups: \nohint, \codeedit, \adaptquiz, and \planquiz. We hosted both study sessions on a web-based platform created specifically for the study. The two sessions were separated by at least a week. As part of the presurvey in Session~I, we collected students' demographic information, prior programming experience, and self-perceived preknowledge. As part of the postsurvey in Session~II, we collected information on their learning experience. 

\looseness-1A total of $398$ students between the ages of $9$--$15$ across $22$ schools participated in the study with the following distribution w.r.t. grade (year of study): $72$ students from grade $4$, $159$ students from grade $5$,  $98$ students from grade $6$, and $69$ students from grade $7$. We refer to students in grades 4--5 as \lowexpstudent{} ($n=231$) and those in grades 6--7 as \highexpstudent{} ($n=167$), and condition our analysis on these categories later. The distribution w.r.t. gender was as follows: $ 195$ females and $203$ males. Distribution of students in each group was as follows: $97$ in \nohint{}, $105$ in \codeedit{}, $97$ in \adaptquiz{}, and $99$ in \planquiz{}.

\subsection{Evaluation Metrics and Analysis Methods}\label{sec:setup.analysis}
After data collection, we processed the following information from each session:
\begin{itemize}
    \item \looseness-1\textit{Session I}: For each \hoc{} task, we analyzed the success rate and time spent on it. Participants could score up to $12$ points ($1$ point per task). We also recorded feedback interaction data per task, including time on the feedback and feedback-seeking rate. Finally, from the presurvey data, we analyzed students' prior programming experience, preknowledge, self-perceived programming interest, and skill.
    \item \textit{Session II}: For each \posttest{} task, we analyzed the success rate. Participants could score up to $15$ points ($1$ point per task). From the postsurvey data, we analyzed students' self-perceived programming interest and skill, as well as their perceived curriculum enjoyment and new skill development. 
\end{itemize}

\looseness-1Since the students' \hoc{} and \posttest{} scores as well as the presurvey and postsurvey data did not follow a normal distribution, we used the \emph{Kruskal-Wallis test}~\cite{ac6c544c-0197-38bd-8c06-ec4f655ff4fd} to evaluate differences across feedback conditions. For all pairwise group comparisons, success rates, time-on-task, feedback time, feedback rate, and Likert-scale survey responses, we applied the \emph{Mann–Whitney U test}~\cite{497e1044-d5b0-30a9-b230-3ca0f10d6f6c}. Since each RQ involves multiple pairwise comparisons across several outcome measures, we applied Bonferroni correction~\cite{Dunn01031961} to control the familywise error rate. Effect sizes are reported as Cliff's $\delta$~\cite{Cliff1993DominanceSO} for significant comparisons vs.\ \nohint{}.

\section{Results and Discussion}\label{sec:results}

\begin{figure*}[!t]
    \centering
    \begin{subfigure}[c]{0.99\textwidth}
    \centering
    \setlength\tabcolsep{5pt}
    \renewcommand{\arraystretch}{1}
    \scalebox{0.95}{
    \begin{tabular}{l cll ll}
        \toprule
        {} & \multicolumn{5}{c}{\textbf{Performance (Success rate \%)}} \\
        \cmidrule(lr){2-6} 
        {} &
        \multicolumn{1}{c}{\small{\textbf{Overall}}} &
        \multicolumn{2}{c}{\textbf{Year of Study (YoS)}} &
        \multicolumn{2}{c}{\textbf{Task Difficulty}} \\
        \cmidrule(lr){2-2}  
        \cmidrule(lr){3-4} \cmidrule(lr){5-6}
        \small{\textbf{Group}} & \small{\hoc{} tasks, all students} & 
        \small{\lowexpstudent{}} & \small{\highexpstudent{}} &
        \small{\learningeasy} & \small{\learninghard} \\
        
        \midrule
        \textbf{\small{\nohint}}  & 
        \cellcolor{gray!0}$73.0 \, \textcolor{gray}{(2.31)}^{\textcolor{white}{\ast\ast}}$ & 
        $ 70.6 \, \textcolor{gray}{(3.11)}$ & 
        $ 75.9 \, \textcolor{gray}{(3.44)}$ &
        \cellcolor{gray!0}$99.2 \, \textcolor{gray}{(0.57)}$ & 
        \cellcolor{gray!0}$59.9 \, \textcolor{gray}{(3.39)}$ \\
        \textbf{\small{\codeedit}}  &  
        \cellcolor{customgreen1!71}$97.2 \, \textcolor{gray}{(1.16)}^{\textcolor{blue}{\ast\ast}}$ & 
         \cellcolor{customgreen1!71}$ 96.9 \, \textcolor{gray}{(1.68)}^{\textcolor{blue}{\ast\ast}}$ & 
         \cellcolor{customgreen1!75}$ 97.7 \, \textcolor{gray}{(1.15)}^{\textcolor{blue}{\ast\ast}}$ & 
        $99.0 \, \textcolor{gray}{(0.95)}$ & 
        \cellcolor{customgreen1!90}$96.3 \, \textcolor{gray}{(1.38)}^{\textcolor{blue}{\ast\ast}}$ \\
        \textbf{\small{\adaptquiz}}  &  
        \cellcolor{customgreen1!42}$86.5 \, \textcolor{gray}{(2.32)}^{\textcolor{blue}{\ast\ast}}$ & 
        \cellcolor{customgreen1!42}$ 86.2 \, \textcolor{gray}{(3.13)}^{\textcolor{blue}{\ast}}$ & 
        \cellcolor{customgreen1!42}$ 86.9 \, \textcolor{gray}{(3.48)}$ & 
        $97.9 \, \textcolor{gray}{(1.14)}$ & 
        \cellcolor{customgreen1!63}$80.8 \, \textcolor{gray}{(3.22)}^{\textcolor{blue}{\ast\ast}}$ \\
        \textbf{\small{\planquiz}}  &  
        \cellcolor{customgreen1!37}$86.2 \, \textcolor{gray}{(2.29)}^{\textcolor{blue}{\ast\ast}}$ & 
         \cellcolor{customgreen1!37}$ 83.8 \, \textcolor{gray}{(3.27)^{\textcolor{blue}{\ast}}}$ & 
        \cellcolor{customgreen1!45}$ 89.5 \, \textcolor{gray}{(3.03)}^{\textcolor{blue}{\ast}}$ & 
        $99.2 \, \textcolor{gray}{(0.76)}$ & 
        \cellcolor{customgreen1!55}$79.7 \, \textcolor{gray}{(3.32)}^{\textcolor{blue}{\ast\ast}}$ \\
        \bottomrule
    \end{tabular}
    }
    \caption{Success rate}
    \label{fig.results.learning.success}
     \end{subfigure}
     \\
     \vspace{1mm}
    \begin{subfigure}[c]{0.99\textwidth}
    \centering
    \setlength\tabcolsep{3pt}
    \renewcommand{\arraystretch}{1}
    \scalebox{0.95}{
    \begin{tabular}{l
        l@{\,$|$\,}r@{\,$|$\,}l
        @{\hspace{14pt}}l@{\,$|$\,}r@{\,$|$\,}r
        @{\hspace{14pt}}l@{\,$|$\,}r@{\,$|$\,}r
        @{\hspace{14pt}}l@{\,$|$\,}r@{\,$|$\,}r
        @{\hspace{14pt}}l@{\,$|$\,}r@{\,$|$\,}r
    }
        \toprule
        {} & \multicolumn{15}{c}{\textbf{\hoc{} task time (s)$|$ Feedback time (s)$|$ Feedback rate}} \\
        \cmidrule(lr){2-16}
        {} &
        \multicolumn{3}{c}{\small{\textbf{Overall}}} &
        \multicolumn{6}{c}{\textbf{Year of Study (YoS)}} &
        \multicolumn{6}{c}{\textbf{Task Difficulty}} \\
        \cmidrule(lr){2-4}
        \cmidrule(lr){5-10}
        \cmidrule(lr){11-16}
        \small{\textbf{Group}} &
        \multicolumn{3}{c}{\small{\hoc{} tasks, all students}} &
        \multicolumn{3}{c}{\small{\lowexpstudent{}}} &
        \multicolumn{3}{c}{\small{\highexpstudent{}}} &
        \multicolumn{3}{c}{\small{\learningeasy{}}} &
        \multicolumn{3}{c}{\small{\learninghard{}}} \\
        \midrule
        \textbf{\small{\nohint}} &
         \textcolor{white}{0000}$351^{\textcolor{white}{\ast\ast}}$ & {-} & {\textcolor{white}{0.0}-} &
        340 & {-} & {-} &
        365 & {-} & {-} &
        223 & {-} & {-} &
        415 & {-} & {-} \\
         \textbf{\small{\codeedit}} &
         \textcolor{white}{0000}$240^{\textcolor{blue}{\ast\ast}}$ & 5 & 0.91 &
        $239^{\textcolor{blue}{\ast\ast}}$ & 5 & 0.90 &
        242 & 5 & 0.92 &
        174 & 2 & 0.26 &
        $273^{\textcolor{blue}{\ast\ast}}$ & 7 & 1.23 \\
        \textbf{\small{\adaptquiz}} &
         \textcolor{white}{0000}$347^{\textcolor{white}{\ast\ast}}$ & 13 & 1.14 &
        355 & 14 & 1.22 &
        338 & 11 & 1.04 &
        273 & 6 & 0.50 &
        384 & 17 & 1.46 \\
        \textbf{\small{\planquiz}} &
         \textcolor{white}{0000}$296^{\textcolor{white}{\ast\ast}}$ & 16 & 1.09 &
        323 & 18 & 1.23 &
        259 & 14 & 0.90 &
        218 & 7 & 0.54 &
        335 & 21 & 1.36 \\
        \bottomrule
    \end{tabular}
    }
    \caption{Time on write-code tasks from \hoc{} and feedback interaction data}
    \label{fig.results.learning.time}
    \end{subfigure}
      \caption{\looseness-1Performance in Session I. \textbf{(a)} presents average success rates on all \hoc{} tasks, also by students' year of study, and by task difficulty. Cell colors indicate deviation from \nohint{} baseline: \textcolor{customgreen1}{green} for improvement, and \textcolor{customred1}{red} for decline, with darker shades for greater differences. All values are reported as means with standard errors in \textcolor{gray}{gray} parentheses. $\textcolor{blue}{\ast}$ indicates significance w.r.t. \nohint{} with $p < \frac{0.05}{30}$, and $\textcolor{blue}{\ast\ast}$ with $p < \frac{0.01}{30}$ (Bonferroni factor is $30$ based on $6$ pairwise comparisons $\times$ $5$ columns). \textbf{(b)} presents average time per \hoc{} task, feedback time, and feedback-seeking rate. $\textcolor{blue}{\ast}$ indicates significance w.r.t. \nohint{} with $p < \frac{0.05}{90}$, and $\textcolor{blue}{\ast\ast}$ with $p < \frac{0.01}{90}$ (Bonferroni factor is $90$ based on $6$ pairwise comparisons $\times$ $15$ columns).}
    \label{fig:results.learning}
\end{figure*}

\subsection{RQ1: Performance in Learning Phase}\label{sec:results.rq1}
\looseness-1To address RQ1, we evaluate the effectiveness of the different feedback methods by comparing their performance on \hoc{} tasks in Session I. The results are presented in Figure~\ref{fig.results.learning.success}. We analyze the average success rate across all $12$ \hoc{} tasks for each group, as well as their performance by student and task difficulty category. All three feedback groups significantly outperformed \nohint{} on \hoc{} success rates.\footnote{Cliff's $\delta$ w.r.t. \nohint{}: \codeedit{} ($0.66$), \adaptquiz{} ($0.37$), \planquiz{} ($0.37$).} \lowexpstudent{} and \highexpstudent{} categories showed trends broadly consistent with the overall results. \learningeasy{} tasks showed a ceiling effect across all groups, with no significant differences. On \learninghard{} tasks, all feedback groups significantly outperformed \nohint{}.

Next, we analyze the time-on-task, feedback time, and feedback-seeking rate. The results are presented in Figure~\ref{fig.results.learning.time}. \codeedit{} significantly reduced time-on-task compared to \nohint{}, while \adaptquiz{} and \planquiz{} did not differ significantly. Feedback time was also lower for \codeedit{} than for \adaptquiz{} and \planquiz{}. The feedback-seeking rate followed a similar pattern, with \codeedit{} seeking feedback less frequently than the quiz-based methods.

\looseness-1These results show that all three feedback methods improve performance during the learning phase. \codeedit{} appears to be the most effective, potentially due to its ease of use and the fact that it directly guides a student to the solution within a few feedback rounds. The quiz-based methods (\adaptquiz{} and \planquiz{}) are associated with more time spent interacting with the feedback. The \learningeasy{} ceiling effect suggests that feedback matters most when tasks are challenging. Both student categories improved consistently across all feedback methods, though \lowexpstudent{} spent longer on quiz-based feedback, either suggesting deeper cognitive engagement or longer time needed to comprehend the quizzes.

\subsection{RQ2: Performance in Post-Learning Phase}\label{sec:results.rq2}
\looseness-1To address RQ2, we evaluate the effectiveness of the different feedback methods by comparing their performance on \posttest{} tasks during Session II. The results are presented in Figure~\ref{fig:results.postlearning}. We first compare the average success rate across all $15$ \posttest{} tasks for each group: no significant differences were observed. These trends remained consistent across \lowexpstudent{} and \highexpstudent{} student categories. \highexpstudent{} students consistently outperformed \lowexpstudent{} students across all feedback groups, reflecting their higher grade levels.

Next, we analyze performance by task category. No significant differences were observed for \postlearningeasy{}, \postlearninghard{}, or \postlearningcommon{} tasks.\footnote{ For \postlearningcommon{} tasks (overlapping with Session~I), Session~I success rates are: \nohint{}: $68.7\ (2.43)$; \codeedit{}: $96.7\ (1.26)$; \adaptquiz{}: $84.7\ (2.40)$; \planquiz{}: $84.3\ (2.50)$. } For the \postlearningnew{} task category involving novel concept combinations, quiz-based methods performed better than both \nohint{} and \codeedit{}, though no individual comparison reached significance. This trend is corroborated by the per-task breakdown in Figure~\ref{fig:posttest_success_per_task} (Appendix~\ref{sec:appendix.results}), where quiz-based groups consistently performed better on P12--P14.

\looseness-1We did not observe any significant differences in success rates in Session~II; notably, differences of feedback groups w.r.t.\ \nohint{} remain positive on all tasks (\posttest{}) and novel tasks (\postlearningnew{}). These results suggest that Session~I feedback did not hinder students' independent problem-solving in Session~II.

\begin{figure}[!t]
	\centering
        \setlength\tabcolsep{5pt}
        \renewcommand{\arraystretch}{1}
    \scalebox{0.95}{
	\begin{tabular}{l c cc cc cc}
	\toprule
        {} & \multicolumn{7}{c}{\textbf{Performance (Success Rate \%)}}
        \\
        \cmidrule(lr){2-8}
		{} & \multicolumn{1}{c}{\textbf{Overall}} & \multicolumn{2}{c}{\textbf{Year of Study (YoS)}} &
        \multicolumn{2}{c}{\textbf{Task Difficulty}} &
        \multicolumn{2}{c}{\textbf{Novelty w.r.t \hoc{}}}
        \\
		\cmidrule(lr){2-2} \cmidrule(lr){3-4}
        \cmidrule(lr){5-6} \cmidrule(lr){7-8}
		\small{\textbf{Group}} & \multicolumn{1}{c}{\small{\posttest}} & \multicolumn{1}{c}{\small{\lowexpstudent{}}} & \multicolumn{1}{c}{\small{\highexpstudent{}}} &
        \multicolumn{1}{c}{\small{\postlearningeasy{}}} &
        \multicolumn{1}{c}{\small{\postlearninghard{}}} &
        \multicolumn{1}{c}{\small{\postlearningcommon{}}} &
        \multicolumn{1}{c}{\small{\postlearningnew{}}}
        \\
		\midrule
		\textbf{\small{\nohint}} & $ 61.1 \, \textcolor{gray}{(2.29)}$ &
         $ 53.6 \, \textcolor{gray}{(2.58)}$ & 
        $ 70.2 \, \textcolor{gray}{(3.55)}$ &
        $ 95.5 \, \textcolor{gray}{(1.16)}$ & 
        $ 52.5 \, \textcolor{gray}{(2.72)}$ &
        $ 79.0 \, \textcolor{gray}{(2.02)}$ & 
        $ 26.1 \, \textcolor{gray}{(3.76)}$ 
        \\
		\textbf{\small{\codeedit}} & $ 63.3 \, \textcolor{gray}{(2.36)}$ & 
         \cellcolor{customgreen1!10}$ 58.0 \, \textcolor{gray}{(2.73)}$ & \cellcolor{customgreen1!10}$ 73.0 \, \textcolor{gray}{(4.03)}$ &
        $ 94.0 \, \textcolor{gray}{(1.26)}$ &
        \cellcolor{customgreen1!10}$ 55.6 \, \textcolor{gray}{(2.76)}$ &
        \cellcolor{customgreen1!10}$ 82.4 \, \textcolor{gray}{(1.94)}$ & 
        \cellcolor{customgreen1!10}$ 30.8 \, \textcolor{gray}{(3.95)}$ 
         \\
		\textbf{\small{\adaptquiz}} & $ 62.3 \, \textcolor{gray}{(2.71)}$ &  
        \cellcolor{customgreen1!5}$ 56.0 \, \textcolor{gray}{(3.42)}$ & 
        $ 70.0 \, \textcolor{gray}{(4.06)}$  &
        \cellcolor{customred1!10}$ 93.1 \, \textcolor{gray}{(1.62)}$ & 
        $ 54.6 \, \textcolor{gray}{(3.14)}$ &
        $ 79.2 \, \textcolor{gray}{(2.34)}$ & 
        \cellcolor{customgreen1!40}$ 35.7 \, \textcolor{gray}{(4.38)}$
        \\
		\textbf{\small{\planquiz}} & $ 62.8 \, \textcolor{gray}{(2.65)}$ & 
        \cellcolor{customgreen1!5}$ 58.0 \, \textcolor{gray}{(3.21)}$ & \cellcolor{customred1!10}$ 69.2 \, \textcolor{gray}{(4.35)}$ &
        \cellcolor{customred1!10}$ 92.9 \, \textcolor{gray}{(1.68)}$ & 
        $ 55.2 \, \textcolor{gray}{(3.06)}$ &
        $ 78.3 \, \textcolor{gray}{(2.35)}$ & \cellcolor{customgreen1!40}$ 36.0 \, \textcolor{gray}{(4.14)}$
        \\
		\bottomrule
	\end{tabular}
    }
    \vspace{-2mm}
	\caption{\looseness-1Performance in Session II. We present average success rates on \posttest{} tasks. Reporting conventions follow Figure~\ref{fig:results.learning}. The Bonferroni factor is $42$ based on $6$ pairwise comparisons $\times 7$ columns; no pairwise comparison reached significance under threshold $p < \frac{0.05}{42}$.
}
	\label{fig:results.postlearning}
\end{figure}

\begin{figure*}[t]
\centering
    \begin{subfigure}[c]{0.53\textwidth}
       \centering
      \includegraphics[width=\textwidth]{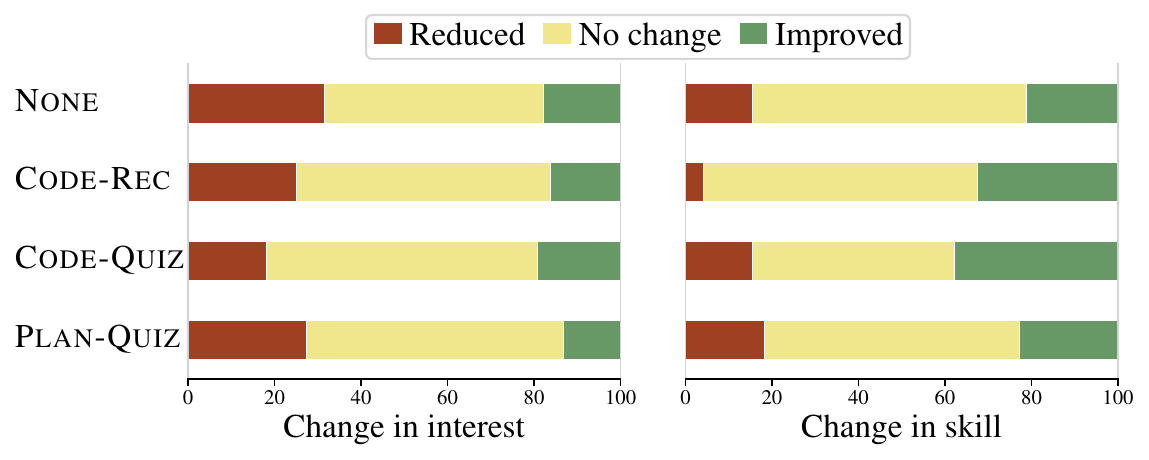}
        \caption{Change in programming interest and skill}
        \label{fig.demographics.skillchange.post}
    \end{subfigure}
     \begin{subfigure}[c]{0.46\textwidth}
        \centering
       \includegraphics[width=\textwidth]{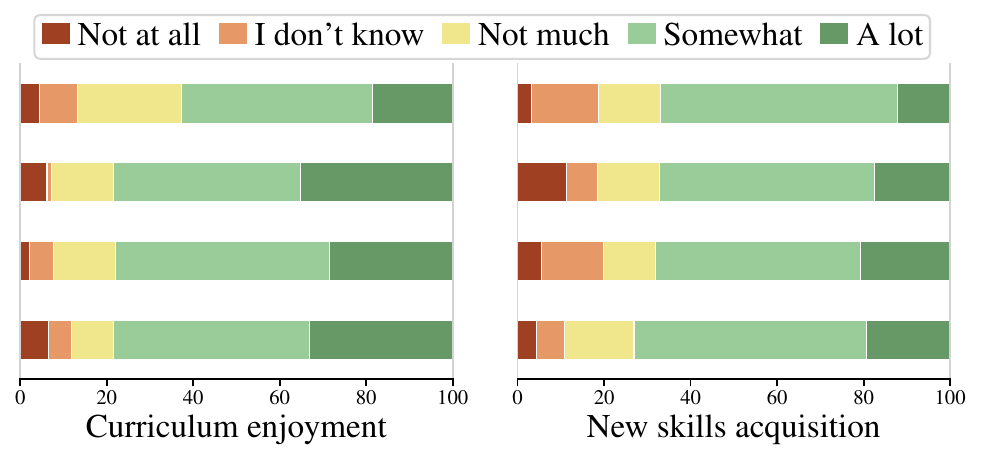}
        \caption{Enjoyment and new skills acquired}
        \label{fig.demographics.enjoyment.post}
    \end{subfigure}
    \vspace{-2mm}
    \caption{\looseness-1Student engagement results based on presurvey and postsurvey responses.}
    \label{fig:demographics.post}
\end{figure*}

\subsection{RQ3: Students' Perceptions}\label{sec:results.rq3}
\looseness-1To address RQ3, we analyze students' perceptions across the different feedback groups, based on presurvey and postsurvey responses on a $5$-point Likert scale. We first examine students' perceived change in programming skill and interest, measured as the difference between pre- and postsurvey responses. The results are presented in Figure~\ref{fig.demographics.skillchange.post}. The \adaptquiz{} group reported a slightly greater increase in both compared to the other groups.

\looseness-1Next, we examine curriculum enjoyment and new skills acquisition. Specifically in the postsurvey, students were asked to rate their curriculum enjoyment (``Did you enjoy solving tasks in the previous session?'') and new skills acquisition (``Did you learn new programming skills by solving tasks in the previous session?'') on a $5$-point Likert scale. The results are presented in Figure~\ref{fig.demographics.enjoyment.post}. We treat ``somewhat'' and ``a lot'' responses as positive indicators. All feedback groups were perceived more positively than \nohint{} for curriculum enjoyment, with positive response rates as follows: \nohint{} ($62.9\%$), \codeedit{} ($78.1\%$), \adaptquiz{} ($76.3\%$), and \planquiz{} ($75.7\%$). This effect was more pronounced among \lowexpstudent{} students, as shown in Figure~\ref{fig:rq3.results.yos} (Appendix~\ref{sec:appendix.results}). For new skills acquisition, the positive response rates were as follows: \nohint{} ($66\%$), \codeedit{} ($68.5\%$), \adaptquiz{} ($66\%$), and \planquiz{} ($73.7\%$), with \planquiz{} showing the highest rating.

\looseness-1All students responded positively to the feedback methods. Notably, the enjoyment boost from quiz-based methods was most pronounced among \lowexpstudent{} students, suggesting they may be particularly engaging for younger learners.

\section{Limitations}\label{limitations}
\looseness-1Next, we discuss a few limitations of our study. First, the focus on K--$8$ learners within a limited geographical region may restrict the generalizability of our findings to other educational contexts. Second, our evaluation of performance is based primarily on task success rates and self-reported perceptions; additional measures of conceptual understanding or long-term retention would provide a more comprehensive view of learning across feedback methods. Third, time-on-task serves as a proxy for cognitive effort, though it may also reflect confounds such as confusion or distraction; think-aloud studies with students would strengthen these interpretations and provide process-level insight into how students engage with the feedback. Finally, while quiz-based methods show a promising trend on post-learning tasks, the group differences are modest and rely on a small number of genuinely novel tasks, so conclusions about transfer effects should be interpreted with caution.


\section{Conclusions}
\label{sec:conclusion}
\looseness-1We conducted a large-scale study comparing three feedback methods for elementary-level visual programming: code-edit recommendations, quizzes based on code edits, and quizzes based on metacognitive strategies focusing on task planning and solution-finding. Our findings have several implications for future research. First, the initial promising results of quiz-based methods on novel concept combinations could be further explored through longitudinal studies examining long-term exposure to these feedback methods. Second, the stronger enjoyment reported by younger students for quiz-based methods motivates further investigation into how feedback design interacts with learner experience across grade levels. Finally, think-aloud protocols would provide process-level insights into how students engage with each method, clarifying the cognitive effects of different feedback approaches.


\appendix

\section{Implementation Details of Feedback Methods}
\label{sec:appendix}

\begin{figure*}[!h]
\centering
    \begin{subfigure}[c]{0.25\textwidth}
        \centering
        \fcolorbox{lightgray}{white}{\includegraphics[trim=0.5cm 0.5cm 0.5cm 0.5cm, clip, height=3.5cm]{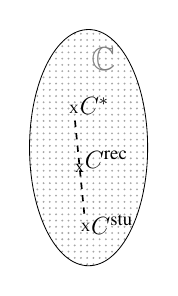}}
        \caption{\codeedit{}}
        \label{fig3.codeonly}
    \end{subfigure}
    \begin{subfigure}[c]{0.35\textwidth}
       \centering
       \fcolorbox{lightgray}{white}{\includegraphics[trim=0.5cm 0.5cm 1.25cm 0.5cm, clip, height=3.5cm]{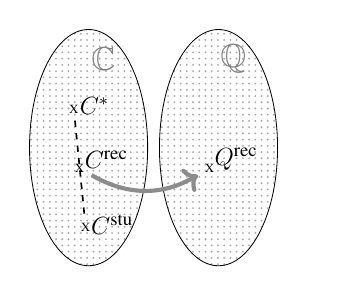}}
        \caption{\adaptquiz{}}
        \label{fig3.codequiz}
    \end{subfigure}
    \begin{subfigure}[c]{0.35\textwidth}
       \centering
       \fcolorbox{lightgray}{white}{\includegraphics[trim=0.5cm 0.5cm 1.25cm 0.5cm, clip, height=3.5cm]{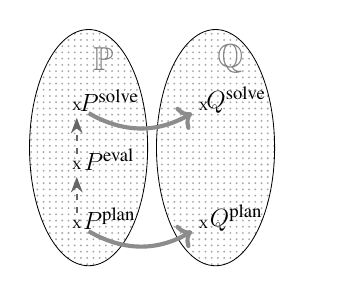}}
        \caption{\planquiz{}}
        \label{fig3.processquiz}
    \end{subfigure}
    \caption{\looseness-1Feedback across code ($\mathbb{C}$), quiz ($\mathbb{Q}$), and metacognitive ($\mathbb{P}$) process spaces.}
    \label{fig3:method}
\end{figure*}

\looseness-1\textbf{Code recommendation based on code-edit (\codeedit{}).} Our \codeedit{} method is inspired by next-step code edit algorithms~\cite{Paassen_Hammer_Price_Barnes_Gross_Pinkwart_2018,DBLP:conf/lats/PiechSHG15,DBLP:conf/its/RiversK14}, and closely follows the implementation in our prior work~\cite{DBLP:conf/aied/GhoshTDS22}. Given the solution code of the write-code task, \solcode{}, and the student's current attempt \stucode{}, the algorithm iteratively expands a neighborhood $N(\text{\stucode})$ of code variations derived from the abstract syntax tree (AST) of \stucode{} until an overlap is found with the rooted subtrees of \solcode{}. The overlapping candidate with the minimal (Zhang-Sasha) tree edit distance~\cite{DBLP:journals/siamcomp/ZhangS89} to \solcode{} is selected as the final code recommendation \codehint{}; see Figure~\ref{fig3.codeonly}.

\looseness-1\textbf{Quiz-based code edit (\adaptquiz{}).} Our implementation of \adaptquiz{} is based on the \textsc{PQuizSyn} algorithm from our prior work~\cite{DBLP:conf/aied/GhoshTDS22}, which builds on the code recommendation generated using the \codeedit{} algorithm. Specifically, to transform the code recommendation into a quiz, we first generate a set of candidate visual task grids using symbolic execution techniques~\cite{DBLP:conf/nips/AhmedCEFGRS20,DBLP:conf/aied/GhoshTDS22}, from which the best visual grid is manually selected. Following~\cite{DBLP:conf/aied/GhoshTDS22}, the missing code-blank in the code recommendation corresponds to the basic action located in the deepest and rightmost child node of the AST. The answer choices for all quizzes are fixed to the three basic actions in the maze domain: \move{}, \turnL{}, and \turnR{}. Through this process, \codehint{} is converted into \quizadapt{}; see Figure~\ref{fig3.codequiz}.

\looseness-1\textbf{Quiz based on metacognitive scaffolds (\planquiz{}).} Our implementation of \planquiz{} involved manually crafting both the planning and solution-finding quizzes for each of the $12$ tasks in the \hoc{} learning phase curriculum. In total, we created $24$ quizzes: $12$ based on the planning and evaluation stages of problem-solving (\quizplan{}), and $12$ based on the solution-finding stage (\quizsolve{}); see Figure~\ref{fig3.processquiz}. Each quiz contained four options, with one correct answer, and we also provided hand-crafted textual feedback for each incorrect option. When a student requests feedback for the first time, \quizplan{} is presented, while \quizsolve{} is shown for all subsequent requests.

\section{Additional Details about Curricula and Results}\label{sec:appendix.results}

\begin{figure*}[h!]
        \centering
        \begin{subfigure}[c]{0.95\textwidth}
        \centering
        \scalebox{0.95}{
            \setlength\tabcolsep{2.8pt}
            \renewcommand{\arraystretch}{1}  
         \begin{tabular}{l || l || l}
            \toprule 
            \multicolumn{1}{c ||}{\textbf{Concept}}
            & \multicolumn{1}{c ||}{\textbf{Tasks in \hoc{}}} & \multicolumn{1}{c}{\textbf{Tasks in \posttest{}}} \\ 
            \midrule
            {Basic moves and turns} & {T01} & {P01 (same as T01), P08}  \\
            \hline
            \DSLRepeat\textcode{\{\}} & {T02, T03, T05} & {P02  (same as T03)} \\
            \DSLRepeat\textcode{\{\};}\DSLRepeat\textcode{\{\}} & {T04} & {} \\
            \DSLRepeat\textcode{\{\};}\DSLRepeat\textcode{\{\};}\DSLRepeat\textcode{\{\}} & {} & {P11} \\
            \hline
            \DSLRepeatUntil\textcode{\{\}} & {T06, T07} & {P03, P04  (same as T07), P09} \\
            \hline
            \DSLRepeatUntil\textcode{\{}\DSLIf\textcode{\}} & {T08, T09} & {P05  (same as T08), P10} \\
            \DSLRepeatUntil\textcode{\{}\DSLIfElse\textcode{\}} & {T10, T11} & {P06  (same as T10)} \\
            \hline
             \DSLRepeatUntil\textcode{\{}\DSLIfElse\textcode{\{\DSLIfElse{}\}\}} & {T12} & {P07  (same as T12), P15} \\
            \hline
            \DSLRepeat\textcode{\{\};}\DSLRepeatUntil\textcode{\{\}} & {} & {P12} (new combination) \\
            \DSLRepeat\textcode{\{}\DSLRepeat\textcode{\}} & {} & {P13} (new combination) \\
            \DSLRepeat\textcode{\{}\DSLIf\textcode{\}} & {} & {P14} (new combination) \\
            \bottomrule
            \end{tabular}
        }
    \caption{\looseness-1Write-code tasks categorized by concepts in their solution codes}
    \label{fig:dist.hocposttest}
    \end{subfigure}
    \\
   \begin{subfigure}[c]{0.98\textwidth}
        \centering
       \includegraphics[height=3.5cm,trim=0 0 0 0,clip]{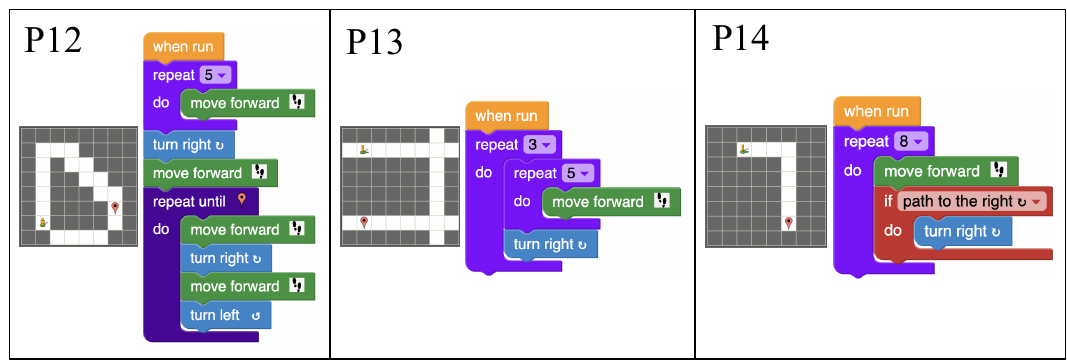}
        \caption{Novel \posttest{} tasks: \postlearningnew{} (P12--P14)}
        \label{fig:tasks.example.posttest} 
    \end{subfigure}
    \caption{\looseness-1Curricula details of the two study sessions. \textbf{(a)} Write-code task distribution by programming concept. \textbf{(b)} \postlearningnew{} tasks with solution codes.}
    \label{fig:curricula_details}
\end{figure*}

Figure~\ref{fig:curricula_details} presents further details of write-code tasks across the two study sessions, including visual task grid and solution code illustrations for \postlearningnew{} tasks.

Figure~\ref{fig:posttest_success_per_task} presents the per-task success rates on \posttest{} tasks for each group. Quiz-based methods consistently perform better on \postlearningnew{} tasks.

\begin{figure*}[!h]
    \centering
    \setlength\tabcolsep{4pt}
    \renewcommand{\arraystretch}{1}
    \scalebox{0.95}{
    \begin{tabular}{l rrccccccccccccc}
        \toprule
        {} & \multicolumn{15}{c}{\textbf{Performance (Success Rate \%)}}
        \\
        \cmidrule{2-16}
        \small{\textbf{Group}} & \small{\textbf{P01}} & \small{\textbf{P02}} & \small{\textbf{P03}} & \small{\textbf{P04}} & \small{\textbf{P05}} & \small{\textbf{P06}} & \small{\textbf{P07}} & \small{\textbf{P08}} & \small{\textbf{P09}} & \small{\textbf{P10}} & \small{\textbf{P11}} & \small{\textbf{P12}} & \small{\textbf{P13}} & \small{\textbf{P14}} & \small{\textbf{P15}}\\
        \midrule
        \textbf{\small{\nohint}} & 
        $100.0$ & $100.0$ & 
        $96.9$ & $96.9$ & 
        $76.3$ & $60.8$ & 
        $40.2$ & $86.6$ & 
        $68.0$ & $38.1$ & 
        $56.7$ & $22.7$ & 
        $34.0$ & $21.6$ & 
        $17.5$ \\
        \textbf{\small{\codeedit}} & $100.0$ & $100.0$ & \cellcolor{customred1!13}$94.3$ & \cellcolor{customgreen1!5}$98.1$ & \cellcolor{customgreen1!9}$78.1$ & \cellcolor{customgreen1!24}$65.7$ & \cellcolor{customgreen1!60}$52.7$ & \cellcolor{customred1!23}$81.9$ & \cellcolor{customred1!21}$63.8$ & \cellcolor{customgreen1!18}$41.9$ & \cellcolor{customgreen1!2}$57.1$ & \cellcolor{customgreen1!34}$29.5$ & \cellcolor{customgreen1!25}$39.0$ & \cellcolor{customgreen1!10}$23.8$ & \cellcolor{customgreen1!31}$23.8$ \\
        \textbf{\small{\adaptquiz}} & \cellcolor{customred1!5}$99.0$ & \cellcolor{customred1!10}$97.9$ & \cellcolor{customred1!30}$90.7$ & \cellcolor{customred1!25}$91.8$ & \cellcolor{customred1!25}$71.1$ & \cellcolor{customgreen1!20}$64.9$ & \cellcolor{customgreen1!51}$50.5$ & \cellcolor{customred1!20}$82.5$ & \cellcolor{customred1!46}$58.8$ & \cellcolor{customred1!10}$36.1$ & \cellcolor{customgreen1!20}$60.8$ & \cellcolor{customgreen1!77}$38.1$ & \cellcolor{customgreen1!36}$41.2$ & \cellcolor{customgreen1!30}$27.8$ & \cellcolor{customgreen1!30}$23.7$ \\
        \textbf{\small{\planquiz}} & \cellcolor{customred1!5}$99.0$ & \cellcolor{customred1!15}$97.0$ & \cellcolor{customred1!19}$92.9$ & \cellcolor{customred1!29}$90.9$ & \cellcolor{customred1!17}$72.7$ & \cellcolor{customgreen1!19}$64.6$ & \cellcolor{customgreen1!26}$45.5$ & \cellcolor{customred1!18}$82.8$ & \cellcolor{customred1!47}$58.6$ & \cellcolor{customgreen1!11}$40.4$ & \cellcolor{customgreen1!24}$61.6$ & \cellcolor{customgreen1!63}$35.4$ & \cellcolor{customgreen1!62}$46.5$ & \cellcolor{customgreen1!23}$26.3$ & \cellcolor{customgreen1!48}$27.3$ \\
        \bottomrule
    \end{tabular}
    }
    \caption{\looseness-1Success rate per \posttest{} task; cell color convention follows Figure~\ref{fig:results.postlearning}.}
    \label{fig:posttest_success_per_task}
\end{figure*}

\begin{figure*}[!h]
\centering
      \includegraphics[width=\textwidth]{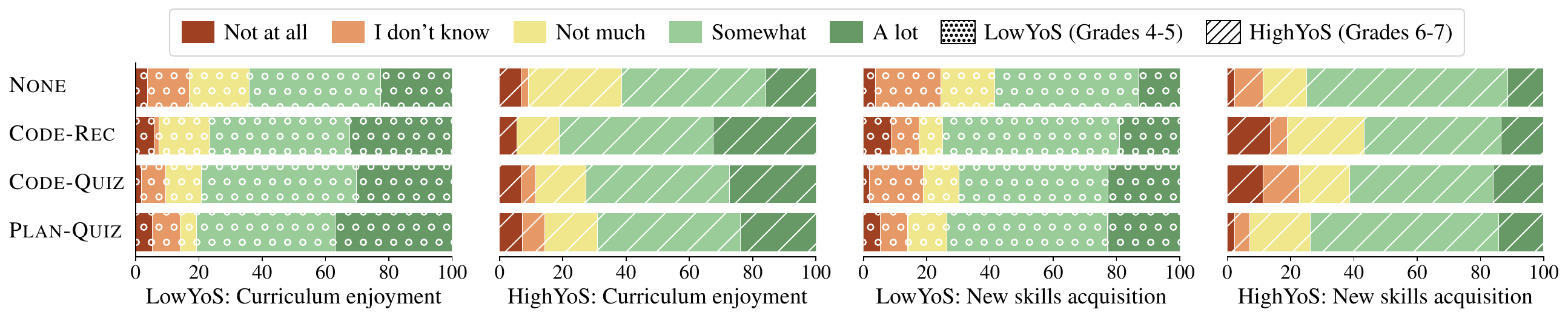}
    \caption{Student engagement results conditioned on their year of study.}
    \label{fig:rq3.results.yos}
\end{figure*}

Figure~\ref{fig:rq3.results.yos} presents perceptions of enjoyment and skill gain per student category. Quiz-based methods showed higher enjoyment among \lowexpstudent{} students.

\begin{acks}
We would like to thank the teachers and students in Estonia for their participation in the study. Funded/Cofunded by the European Union (ERC, TOPS, 101039090). Views and opinions expressed are however those of the author(s) only and do not necessarily reflect those of the European Union or the European Research Council. Neither the European Union nor the granting authority can be held responsible for them.
\end{acks}

\bibliographystyle{ACM-Reference-Format}
\bibliography{main}

\end{document}